\documentclass[10pt,onecolumn,letterpaper]{article}

\usepackage{physics}
\usepackage{graphicx}
\usepackage{subcaption}
\usepackage[pagenumbers]{cvpr} 

\newcommand{\nNodes}{n_{\text{nodes}}}
\newcommand{\Tdepol}{T_{\text{depol}}}
\newcommand{\Ftarget}{F_{\text{target}}}
\newcommand{\Npur}{N_{\text{pur}}}
\newcommand{\psucc}{p_{\text{succ}}}
\newcommand{\Latt}{L_{\text{att}}}

\definecolor{cvprblue}{rgb}{0.21,0.49,0.74}
\usepackage[pagebackref,breaklinks,colorlinks,allcolors=cvprblue]{hyperref}

\def\confName{COMPSCI690@UMass}
\def\confYear{2025}

\title{Simulation of Quantum Repeater Networks under Decoherence and Purification Constraints}

\author{Wenhan Li\\
{\tt\small wenhanli@umass.edu}
\and
Shiyu Zhang\\
{\tt\small shiyuzhang@umass.edu}
}

\begin{document}
\maketitle
\begin{abstract}

Long-distance quantum communication requires reliable entanglement distribution, but direct generation with protocols such as Barrett--Kok suffers from exponentially decreasing success probability with distance, making it impractical over hundreds of kilometers. Quantum repeaters address this by segmenting the channel and combining entanglement generation, swapping, and purification. In this work, we present a simulation framework for chain-based repeaters under continuous-time depolarizing noise. Our model implements heralded entanglement generation, Bell-state swapping, and multi-round purification, with configurable chain length, noise levels, and purification depth. Numerical results highlight how memory decoherence constrains performance, how purification mitigates fidelity loss, and how time and entanglement costs scale with distance. While simplified, the framework offers a flexible tool for exploring trade-offs in repeater design and provides a basis for extensions toward more complex network scenarios.
\end{abstract}

\section{Introduction}
Quantum communication is a critical building block for the emerging vision of the quantum internet, enabling secure key distribution, distributed quantum computing, and large-scale quantum networks. However, establishing high-fidelity entanglement across long distances is challenging due to optical channel loss and memory decoherence. Heralded entanglement generation schemes such as the Barrett--Kok (BK) protocol~\cite{Barrett_2005} suffer from exponentially decreasing success probability with distance, limiting direct entanglement to tens of kilometers.\\

Quantum repeaters overcome this limitation by dividing the total distance into shorter segments~\cite{Briegel_1998, Dr1999}. Entanglement is first generated locally, then extended via entanglement swapping, and stabilized by purification~\cite{Bennett_1996}. This approach mitigates the exponential decay of BK success but introduces new costs: each swap reduces fidelity, idle qubits decohere, and purification consumes additional entangled pairs and time.\\

Purification protocols such as BBPSSW~\cite{Bennett_1996} can improve fidelity but require many resources. While more advanced approaches, including error-corrected and all-photonic repeaters~\cite{Muralidharan2016,Azuma_2023,sangouard2009quantumrepeatersbasedatomic}, promise better scalability, they remain experimentally challenging. Consequently, purification-driven repeaters are still the most practical near-term approach. At the same time, purification alone is not scalable to very long chains. Our work therefore uses purification as a reference point to study cost and performance trade-offs. By quantifying the Werner-pair consumption and latency under realistic noise, we show why purification becomes a bottleneck and highlight the motivation for future exploration of quantum error correction and more advanced architectures.\\

This study provides a flexible simulation framework to explore these effects. We implement BK entanglement generation, multi-round purification, entanglement swapping, and continuous-time depolarization, with configurable parameters. The simulator focuses on quantifying the resource and time overhead needed to meet a target fidelity. Our contributions are: 
\begin{itemize}
    \item A configurable, modular simulator for repeater chains with noise and purification.
    \item Quantitative analysis of decoherence impacts and purification costs.
    \item A foundation for future work toward QEC-based or hybrid repeater designs.
\end{itemize}

\section{Modeling Assumptions and Implementation}

\begin{table}[h]
\centering
\caption{Simulation parameters and notation}
\begin{tabular}{ll}
\hline
Symbol & Description \\
\hline
$L$ & Total communication distance (km) \\
$\nNodes$ & Number of repeater nodes \\
$\Tdepol$ & Memory depolarization time constant (ms) \\
$\Ftarget$ & Target end-to-end fidelity \\
$\Npur$ & Number of purification rounds \\
$\psucc$ & Success probability of BK entanglement generation \\
\hline
\end{tabular}
\end{table}

\subsection{Atomic Entanglement Generation}

Entanglement between adjacent nodes is established using the Barrett--Kok (BK) protocol~\cite{Barrett_2005}. 
Two users are connected by an optical fiber with a heralding station placed at the midpoint. 
Each photon travels a distance of $L/2$ at speed $c \approx 2\times 10^{8}~\text{m/s}$, requiring $(L/2)/c$ time. 
Because heralding requires a round-trip classical signal, the total latency per attempt is
\begin{equation}
    \tau = \frac{L}{c},
\end{equation}
where $L$ is the user-to-user separation. The fiber transmissivity is given by
\begin{equation}
    \eta = e^{-L/\Latt},
\end{equation}
with attenuation length $\Latt = 22.5~\text{km}$. 
The per-attempt success probability of entanglement generation is then
\begin{equation}
    \psucc = \frac{\eta^2}{2}.
\end{equation}

The number of attempts until success is modeled as a geometric random variable with parameter $\psucc$. Multiplying the sampled count by $\tau$ gives the total generation time for an elementary link.

\subsection{Network Simulation}

To reduce the exponential latency of direct BK entanglement, we place $N_r$ quantum repeaters equally spaced between the end users Alice and Bob. 
The network is represented as a graph with nodes connected by weighted edges, and all quantum state evolution (unitaries and measurements) is simulated using QuTiP. \\

Each repeater executes:
\begin{enumerate}
  \item Initialize two memory qubits, one for entanglement with the left neighbor and one with the right.
  \item Perform a Bell‐state measurement on the two internal qubits.
  \item Transmit the measurement outcome (classical bits) to the designated neighbor, who applies the corresponding Pauli correction.
\end{enumerate} 
Steps (2)–(3) require time $o+\tau$, where $o$ is the local-operation latency and $\tau$ is the classical-communication delay. 
Disjoint repeaters can perform these operations in parallel. \\

Entangled links are represented as objects storing the node pair and the associated $4\times 4$ density matrix. 
Entanglement swapping is simulated by applying the projector
\[
P = I \otimes |\Phi^+\rangle\langle\Phi^+| \otimes I
\]
to the four-qubit state, renormalizing, and tracing out the middle qubits. 
We restrict to the $|\Phi^+\rangle$ outcome since other Bell outcomes differ only by local Pauli corrections.

To generate end-to-end entanglement, BK links are first established between all adjacent node pairs. 
We then perform \emph{parallel} entanglement-swapping rounds: in each round, adjacent pairs are grouped, swapped simultaneously, and replaced by the new links, halving the list size. 
After $K = \lceil \log_2 (N_r+1)\rceil$ rounds, a single Bell pair connects Alice and Bob. 
Since each round takes $o+\tau$ time, the total swap time is
\[
  K\,(o + \tau) = \lceil \log_2 (N_r+1)\rceil\,(o + \tau).
\]

\subsection{Continuous-Time Depolarization Noise}

The sole noise source modeled is continuous-time depolarization in memory qubits. 
For any two-qubit density matrix, evolution is given by
\[
  \rho(t) = e^{-t/\Tdepol}\,\rho(0) + \bigl(1-e^{-t/\Tdepol}\bigr)\,\tfrac{I_4}{4},
\]
where $\Tdepol$ is the memory lifetime and $I_4$ is the $4\times4$ identity matrix.

We assume perfect unitaries and classical communication. 
After a successful herald, memory qubits decohere for $\tau$ ms while waiting for the heralding result. 
Each swap round requires $o+\tau$ ms, during which the entangled state further decoheres. 
No additional noise sources are included.
\subsection{Purification Protocol}

We employ the BBPSSW protocol~\cite{Bennett_1996} to preserve high fidelity during end-to-end entanglement. 
Purification is triggered at the start of each entanglement-swap round whenever the link fidelity falls below a calculated threshold. 
We assume that any two adjacent nodes can prepare an arbitrary number of identical Werner pairs in parallel:
\[
  \rho = \omega\,|\Phi^+\rangle\langle\Phi^+|
        + (1-\omega)\,\tfrac{I_4}{4}, \quad \omega \in [0,1].
\]

Each purification round updates the fidelity $F$ and succeeds with probability $p_{\rm succ}$:
\begin{align}
  F &\;\longrightarrow\;
    \frac{F^2 + \tfrac{1}{9}(1 - F)^2}
         {F^2 + \tfrac{2}{3}\,F(1 - F) + \tfrac{5}{9}(1 - F)^2},
    \label{eq:purified-fidelity}\\[6pt]
  p_{\rm succ}
   &= F^2 + \tfrac{2}{3}\,F(1 - F) + \tfrac{5}{9}(1 - F)^2.
    \label{eq:success-prob}
\end{align}
For a lower-bound estimate, we set $p_{\rm succ}=1$. 
After purification the state is reset to
\[
  \rho' = F'\,|\Phi^+\rangle\langle\Phi^+|
       + \tfrac{1-F'}{3}\bigl(I_4 - |\Phi^+\rangle\langle\Phi^+|\bigr).
\]

To guarantee a target fidelity $F_{\rm target}$ after $K = \lceil \log_2 N_{\rm ent} \rceil$ swap stages, we compute the minimum input fidelity required at each layer. 
A single swap of two Werner states of fidelity $\omega$ produces fidelity
\[
  F' = \frac{1 + 3\,\omega^2}{4}
     = \frac{1 + \tfrac{(4F-1)^2}{3}}{4}.
\]
Requiring $F' > F_{\rm target}$ yields the threshold
\[
  F > \frac{\sqrt{12F_{\rm target}-3} + 1}{4}.
\]
Iterating this inequality backward across the $K$ swap layers gives the minimum pre-swap fidelity required to achieve $F_{\rm target}$. 
Although this derivation neglects depolarization, it provides a lower bound on the number of purification rounds needed, with a corresponding Werner-pair cost scaling as $2^K$.

\subsection{Purification Rounds: Ideal vs. Actual Execution under Decoherence}

We distinguish between two cases.  
In the \emph{ideal} analysis, the number of BBPSSW rounds is estimated assuming (i) no decoherence during operations and (ii) $p_{succ}=1$. 
This yields the theoretical minimum number of rounds required to reach $F_{\rm target}$, based solely on swap-induced fidelity loss.

In the \emph{actual} simulation, however, qubits undergo continuous-time depolarization not only during BK generation but also while waiting in memory throughout purification and swap operations. 
When $T_{{depol}}$ is small, this accumulated noise can reduce fidelity so severely that even the ideal number of purification rounds fails to restore $F_{\rm end}$ to $F_{\rm target}$. 
In our configuration, this breakdown occurs for $T_{{depol}} < 4$ ms, establishing a practical noise-tolerance boundary.

\subsection{Evaluation Metrics}

We evaluate performance using three metrics:
\begin{enumerate}
  \item Entanglement generation time,
  \item End-to-end final fidelity $F_{\rm end}$, and
  \item \emph{Purification cost}, defined as the average number of Werner pairs consumed per node.
\end{enumerate}

To compute the purification cost, let
\[
  \{c_0, c_1, \dots, c_n\}
\]
denote the average number of Werner pairs required at iteration $k$ of the swap protocol ($c_0$ being the cost of producing one raw pair).  
Since each purified pair at layer $k$ consumes $c_{k-1}$ pairs from layer $k-1$, the total number of raw pairs needed to realize $c_k$ purified pairs is
\[
  \prod_{j=0}^{k} c_j.
\]
Summing across all layers $k=0,\dots,n$ yields the total initial cost per node:
\[
  C_{\rm total}
  = \sum_{k=0}^{n} \prod_{j=0}^{k} c_j.
\]
We refer to $C_{\rm total}$ as the \emph{final purification cost}.

\subsection{Implementation and Code Availability}
Our simulator is implemented in Python (QuTiP for state evolution, NumPy for numerics, Matplotlib for plotting). 
Each experiment programmatically composes Barrett–Kok link generation, parallel swapping rounds, 
and BBPSSW purification with continuous-time depolarization. All source code is available at:
\url{https://github.com/OceanCatSZ/Quantum-Repeater-Sim}.

\section{Experiments and Results}

\subsection{Effect of $T_{depol}$ on Performance}

\begin{figure}[h]
    \centering
    \begin{subfigure}[b]{0.32\textwidth}
        \includegraphics[width=\linewidth]{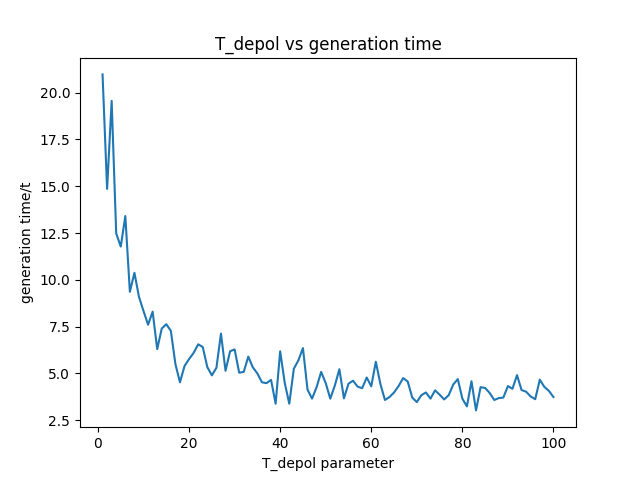}
        \caption{Generation Time}
        \label{fig:generation-time}
    \end{subfigure}
    \hfill
    \begin{subfigure}[b]{0.32\textwidth}
        \includegraphics[width=\linewidth]{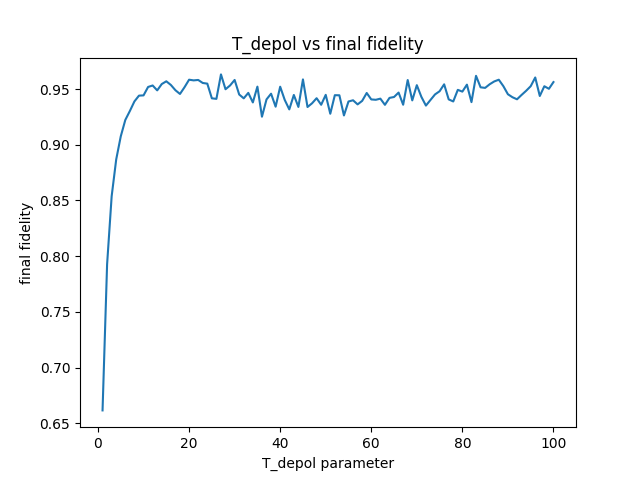}
        \caption{Final Fidelity}
        \label{fig:final-fidelity}
    \end{subfigure}
    \hfill
    \begin{subfigure}[b]{0.32\textwidth}
        \includegraphics[width=\linewidth]{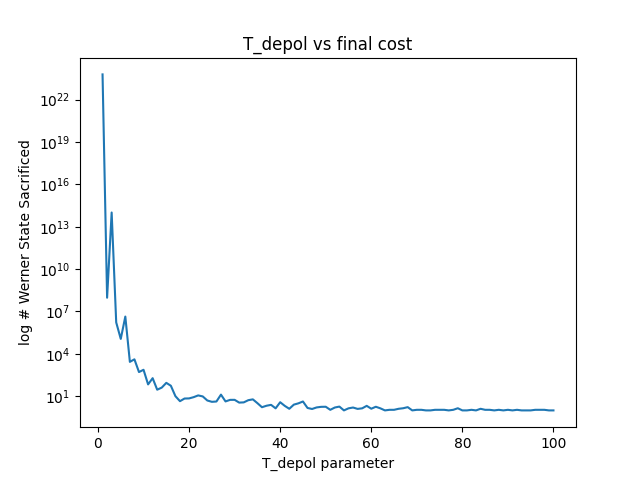}
        \caption{Werner-Pair Cost}
        \label{fig:werner-cost}
    \end{subfigure}
    \caption{Impact of varying $T_{depol}$ on generation time, fidelity, and purification cost.}
    \label{fig:tdepol-summary}
\end{figure}

We first examine the effect of the decoherence time $T_{depol}$ using $L=200$ km between Alice and Bob and $n=6$ repeater nodes.  \\

Figure~\ref{fig:generation-time} shows that generation time decreases as $T_{depol}$ increases. For small $T_{depol}$, entangled states decohere rapidly, necessitating more purification rounds to reach the target fidelity, which in turn increases the total generation time. In contrast, for larger $T_{depol}$, decoherence is weaker, fewer purification rounds are required, and the generation time shortens. \\

Figure~\ref{fig:final-fidelity} shows that the final fidelity improves with larger $T_{depol}$. When $T_{depol}$ is small, waiting and swapping operations cause fidelity to degrade even after purification. For larger $T_{depol}$, decoherence is negligible, allowing fidelity to remain high across the protocol.\\

Finally, Figure~\ref{fig:werner-cost} shows that purification cost grows dramatically as $T_{depol}$ decreases. For short decoherence times, each purification stage consumes many Werner pairs, producing a total cost on the order of $10^{22}$ pairs (logarithmic scale). With larger $T_{depol}$, the purification demand is reduced, significantly reducing the number of Werner states that must be sacrificed.

\subsection{Noise Tolerance Boundary}

We set a target fidelity threshold of $F_{end} \geq 0.9$.  
Our simulations show that this threshold can only be met when $T_{depol} \geq 4$ ms. For smaller values, decoherence dominates, and purification cannot fully recover fidelity. Thus, $T_{depol}=4$ ms defines a practical noise-tolerance boundary for the protocol.

\subsection{Minimum Werner-Pair Cost and Time Efficiency}

For large $T_{depol}$, the number of Werner pairs sacrificed during purification is minimized. In our configuration, the minimum requirement was
\[
  7 \times 4 \times 2 \times 1 = 56 \quad \text{pairs}.
\]
As also shown in Figure~\ref{fig:generation-time}, larger $T_{depol}$ simultaneously reduces both generation time and purification cost. Long coherence times therefore improve both time efficiency and resource efficiency.

\subsection{Influence of Distance on Entanglement Generation}

To study distance effects, we fixed $T_{depol}=10$ ms and varied the number of repeater nodes ($n=4$ to $40$) for three distances: $L=200$, $500$, and $1000$ km.

\begin{figure}[h]
    \centering
    \begin{subfigure}[b]{0.32\textwidth}
        \includegraphics[width=\linewidth]{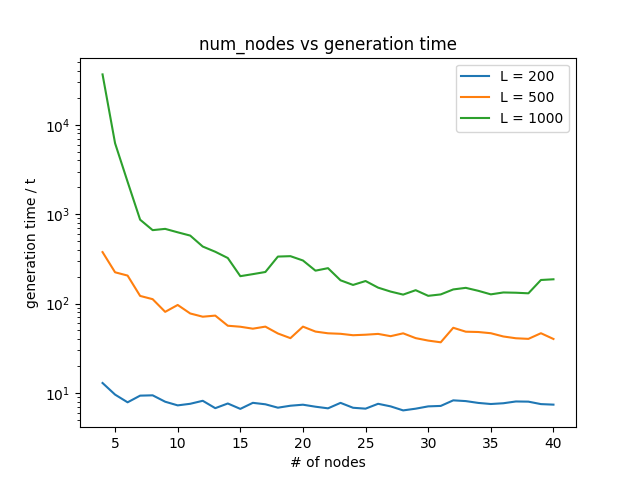}
        \caption{Generation Time}
        \label{fig:distance-time}
    \end{subfigure}
    \hfill
    \begin{subfigure}[b]{0.32\textwidth}
        \includegraphics[width=\linewidth]{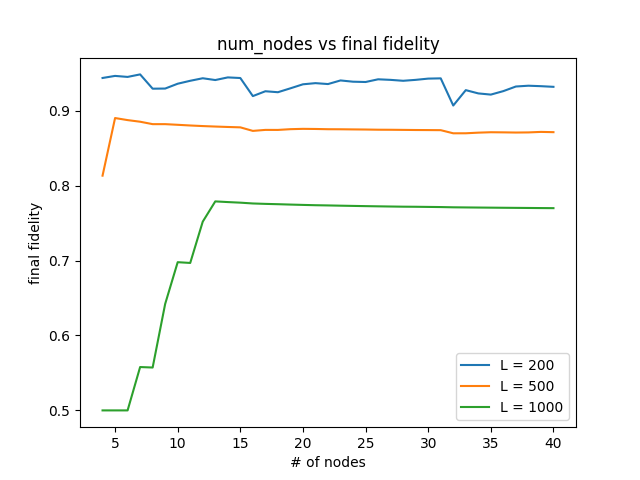}
        \caption{Final Fidelity}
        \label{fig:distance-fidelity}
    \end{subfigure}
    \hfill
    \begin{subfigure}[b]{0.32\textwidth}
        \includegraphics[width=\linewidth]{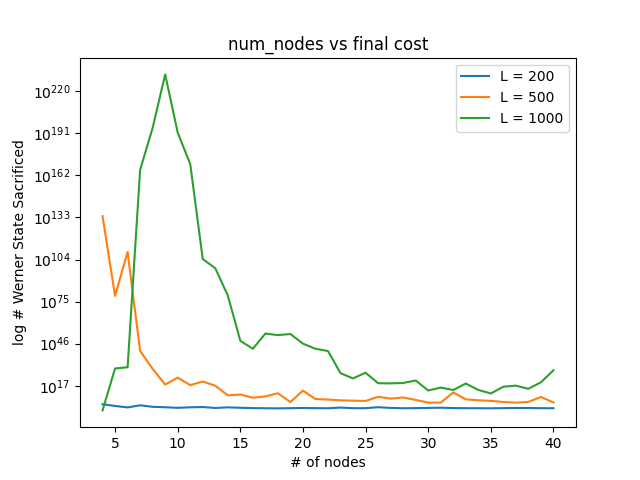}
        \caption{Werner-Pair Cost}
        \label{fig:distance-cost}
    \end{subfigure}
    \caption{Effect of total distance $L$ on entanglement generation performance.}
    \label{fig:distance-effect}
\end{figure}

As shown in Figure~\ref{fig:distance-effect}, increasing $L$ leads to longer generation times, lower final fidelity, and higher Werner-pair costs. Larger distances exacerbate decoherence and photon loss, requiring more purification rounds to maintain target fidelity. Even with repeated purification, fidelity degrades as $L$ increases, and resource consumption rises accordingly.

\subsection{Effect of the Number of Quantum Repeaters}

For each distance $L$, increasing the number of repeater nodes improves performance: generation time decreases and fewer Werner pairs are consumed.  \\

The optimal number of repeaters for maximizing fidelity depends on $L$. For $L=200$ km, the maximum fidelity occurs at $n=6$ nodes; for $L=500$ km, at $n=8$ nodes; and for $L=1000$ km, at $n=13$ nodes.  \\

In general, deploying a moderate number of repeaters is sufficient to achieve high fidelity. However, if the objective is to minimize generation time and resource consumption, adding more repeaters is beneficial.

\section{Conclusion}

In this project, we simulated a quantum network architecture leveraging quantum repeaters to efficiently generate high-fidelity entanglement between two distant parties—Alice and Bob—under realistic physical constraints such as decoherence and limited entanglement success rates. By modeling the Barrett-Kok (BK) protocol for entanglement generation, incorporating continuous-time depolarization as the primary noise model, and applying the BBPSSW purification protocol, we were able to assess the impact of decoherence on fidelity, time, and resource consumption.\\\\
Our experiments demonstrate that the entanglement generation time and purification cost grow rapidly with increased noise, while high decoherence (i.e., low $T_{\text{depol}}$) severely limits the final fidelity. Specifically, we observed that maintaining $T_{\text{depol}} \geq 4$~ms is necessary to achieve end-to-end fidelity above 0.9. Moreover, we quantified the number of Werner states sacrificed in purification, highlighting the trade-off between fidelity and resource efficiency.\\\\
This simulation confirms the critical importance of purification and the strategic placement of quantum repeaters in real-world quantum communication systems. While simplifications were made (e.g., perfect operations and heralding), our model captures key insights into optimizing quantum network performance and serves as a foundation for future extensions involving realistic hardware constraints. However, our results also verified that the purification scheme requires Werner states exponentially increasing with respect to distance, and this will lead us to explore alternative entanglement distribution schemes, and error correction techniques.

\bibliographystyle{unsrt}  
\bibliography{references} 
\end{document}